\documentclass[review]{elsarticle}

\usepackage{lineno,hyperref}

\journal{Astronomy and Computing}

\usepackage{amsmath}
\usepackage{multirow}
\usepackage{float}
\usepackage[dvipsnames]{xcolor}
\usepackage{float}

\usepackage{tikz}
\usetikzlibrary{shapes.geometric, arrows}

\tikzstyle{cosmoparams} = [rectangle, rounded corners, minimum width=3cm, minimum height=1cm,text centered, draw=black, fill=gray!30]
\tikzstyle{bkg} = [rectangle, minimum width=3cm, minimum height=1cm,text centered, draw=black, fill=green!30]
\tikzstyle{pk} = [rectangle, minimum width=3cm, minimum height=1cm,text centered, text width=5.45cm, draw=black, fill=ProcessBlue!30]
\tikzstyle{boltz} = [rectangle, minimum width=3cm, minimum height=1cm, text centered, text width=4.cm, draw=black, fill=OrangeRed!30]
\tikzstyle{cls} = [rectangle, minimum width=3cm, minimum height=1cm,text centered, draw=black, fill=JungleGreen!30]
\tikzstyle{fitfunct} = [rectangle, minimum width=3cm, minimum height=1cm, text centered, text width=1cm, draw=black, fill=ProcessBlue!30]
\tikzstyle{arrow} = [thick,->,>=stealth]

\usepackage{booktabs, ltablex, siunitx, threeparttablex, caption}
\usepackage{xcolor,colortbl}
\usepackage{caption}

\biboptions{authoryear}

\begin{document}
\begin{frontmatter}

\title{Predicting Cosmological Observables with \textsf{PyCosmo}}


\author{F.Tarsitano\fnref{mymainaddress}} 
\author{U.Schmitt\fnref{mymainaddress}}
\author{A.Refregier\fnref{mymainaddress}}
\author{J.Fluri\fnref{mymainaddress}}
\author{R.Sgier\fnref{mymainaddress}}
\author{A.Nicola\fnref{princetonaddress}}
\author{J.Herbel\fnref{mymainaddress}}
\author{A.Amara\fnref{mymainaddress}}
\author{T.Kacprzak\fnref{mymainaddress}}
\author{L.Heisenberg\fnref{theoryaddress}}

\address[mymainaddress]{Institute for Particle Physics and Astrophysics, Department of Physics, ETH Zurich,
	Wolfgang-Pauli-Strasse 27, 8093 Zurich, Switzerland}
\address[theoryaddress]{Institute for Theoretical Physics, ETH Zurich, Wolfgang-Pauli-Strasse 27, 8093, Zurich, Switzerland}
\address[princetonaddress]{Department of Astrophysical Sciences, Princeton University, Princeton, NJ 08544, USA}

\begin{abstract}
Current and upcoming cosmological experiments open a new era of precision cosmology, thus demanding accurate theoretical predictions for cosmological observables. Because of the complexity of the codes delivering such predictions, reaching a high level of numerical accuracy is challenging. Among the codes already fulfilling this task, \textsf{PyCosmo} is a Python-based framework providing solutions to the Einstein-Boltzmann equations and accurate predictions for cosmological observables. In this work, we first describe how the observables are implemented. Then, we check the accuracy of the theoretical predictions for background quantities, power spectra and Limber and beyond-Limber angular power spectra by comparison with other codes: the Core Cosmology Library (\texttt{CCL}), \texttt{CLASS}, \texttt{HMCode} and \texttt{iCosmo}.
In our analysis we quantify the agreement of \textsf{PyCosmo} with the other codes, for a range of cosmological models, monitored through a series of \textit{unit tests}. \textsf{PyCosmo}, conceived as a multi-purpose cosmology calculation tool in \texttt{Python}, is designed to be interactive and user-friendly. A current version of the code (without the Boltzmann Solver) is publicly available and can be used interactively on the platform \textsf{PyCosmo Hub}, all accessible from this link: (\url{https://cosmology.ethz.ch/research/software-lab/PyCosmo.html}). On the hub the users can perform their own computations using \texttt{Jupyter Notebooks} without the need of installing any software, access to the results presented in this work and benefit from tutorial notebooks illustrating the usage of the code. The link above also redirects to the code release and documentation. 
\end{abstract}


\end{frontmatter}


\section{Introduction}\label{introduction}

\noindent Present research in cosmology investigates the validity of the  $\Lambda$CDM model and its extensions by testing its parameters through observational probes, such as the Cosmic Microwave Background (CMB), Baryonic Acoustic Oscillations (BAO), weak lensing, cluster counts, supernovae and galaxy surveys. The combination of these observables has high constraining power on the parameters of these cosmological models. Current and upcoming cosmological experiments, such as DES\footnote{\url{http://www.darkenergysurvey.org}}, DESI\footnote{\url{http://desi.lbl.gov}}, LSST\footnote{\url{http://www.lsst.org}}, Euclid\footnote{\url{http://sci.esa.int/euclid/}} and WFIRST\footnote{\url{http://wfirst.gsfc.nasa.gov}} aim at precise measurements of these observables, thus demanding highly accurate theoretical predictions.
Codes fulfilling this task are already available, such as \texttt{COSMICS} \citep{cosmics1995}, \texttt{CMBFAST} \citep{cmbfast1996}, \texttt{CMBEASY} \citep{cmbeasy2005}, \texttt{CAMB} \citep{camb2000}, \texttt{CLASS} \citep{class2011}, \textsf{iCosmo} \citep{iCosmo2011}, \texttt{CosmoLike} \citep{krause2017}, \texttt{CosmoSIS} \citep{zuntz2015}, \texttt{CCL} \citep{ccl2019}. \texttt{PyCosmo} \citep{Alex2018} is a recently introduced Python-based framework which provides cosmological model predictions, fitting within the upcoming new era of precision cosmology. As a Boltzmann solver, it computes solutions to the set of Einstein-Boltzmann equations, which govern the linear evolution of perturbations in the Universe. These calculations are at the core of most cosmological analyses. \textsf{PyCosmo} introduces a novel architecture that uses symbolic calculations. As described in a previous work \citep{Alex2018} the code, based on the \textit{Python} library \textit{Sympy} \citep{Sympy2017}, uses computer algebra capabilities to produce fast and accurate solutions to the set of Einstein-Boltzmann equations, and provides the user a convenient interface to manipulate the equations and implement new cosmological models. \\
In this paper, we present \textsf{PyCosmo} as a more general cosmology code, providing accurate predictions for cosmological quantities, defined in terms of background computations, linear and non-linear perturbations and observables. The fitting functions for the linear and non-linear power spectrum, which are used to compute predictions for angular power spectra with the Limber Approximation \citep{Limber2008}, have been extensively tested. In particular we refer to the \texttt{Halofit} fitting function \citep{Halofit_Smith2003, Halofit_Takahashi2012} and to a revised version of the Halo Model, presented in \cite{mead2015} as a more accurate function which also accounts for baryonic feedback; below in this section we will refer to it as the \textit{Mead et al. model}. Both fitting functions within \textsf{PyCosmo} have been used in the MCCL analysis of the DES data described in \cite{MCCL2019}. The CMB angular power spectrum is computed using the approach of line-of-sight integration proposed in \cite{LOS1996}. \\
In order to assess the accuracy of such computations it is important to compare \textsf{PyCosmo} to other available codes, with the aim of obtaining the highest possible agreement between algorithms with independent implementations. In \textsf{PyCosmo} such comparisons are constantly monitored through a system of unit tests. 
Conceived as a user-friendly code, the currently tested and validated version of \textsf{PyCosmo} is currently available on a public hub, called \textsf{PyCosmo Hub} and accessible from \url{https://cosmology.ethz.ch/research/software-lab/PyCosmo.html}. This server hosts several \texttt{Jupyter} notebooks showing how to use \textsf{PyCosmo} by including tutorial-notebooks and examples. Registered users can use \textsf{PyCosmo} for their own calculations without the need of local installations. More details about the hub will be provided later in Section \ref{architecture}.\\
This paper focuses on the implementation of the cosmological observables and the tests made in order to check their accuracy. In this context \textsf{PyCosmo} is compared to the following codes (see also \cite{ccl2019} for an earlier comparison of some of these codes):

\begin{itemize}
	\item \texttt{CLASS} \citep{class2011}, a C-based Boltzmann solver widely used to compute theoretical predictions, and its python wrapper, \texttt{classy};
	\item  \texttt{iCosmo} \citep{iCosmo2011}, an earlier cosmology package written in IDL;
	\item Core Cosmology Library \citep{ccl2019}, developed within the LSST Dark Energy Science Collaboration \citep{LSST2012};
	\item \texttt{HMCode} \citep{mead2015}, the original implementation of the \textit{Mead et al. model}, coded in \texttt{Fortran}.
\end{itemize}

\noindent 
In Section \ref{theory} we give an overview of the cosmological observables implemented in \textsf{PyCosmo}. Section \ref{implementation} describes how they are implemented, providing details concerning the code architecture. Information about the \textsf{PyCosmo Hub} is also provided. In Section \ref{validation} we describe the setup and the conventions used for code comparison and we  present the main results from those tests.

\section{Cosmological model} \label{theory}
\noindent
In this section we give definitions for the cosmological models implemented in \textsf{PyCosmo}.
The current version of the code supports a $\Lambda$CDM cosmology, defined in terms of the matter density components $\Omega_b$ and $\Omega_m$, the Hubble parameter $H_0$, spectral index $n_s$, normalization of the density fluctuations $\sigma_8$ and a dark energy model with equation-of-state $w=-1$. The curvature is defined by $\Omega_k = 1-\sum_{i}\Omega_i$, where $i$ refers to matter ($\Omega_m$), radiation ($\Omega_r$) and vacuum ($\Omega_{\Lambda}$) density components. 

\subsection{Background}
\noindent
Background computations start with the calculation of the Hubble parameter, $H(a)$, and the cosmological distances. The basis of such calculations is the Friedmann equation, obtained by applying the Einstein's equations to the FLRW metric:
\begin{equation}
\left(\frac{1}{a}\frac{da}{dt}\right)^{2} = \frac{8\pi G}{3}\rho + \frac{(1-\Omega)H_0^{2}}{a^{2}}.
\label{Friedmann_equation}
\end{equation}

\noindent In this equation $G$ is the Newton's constant, $\rho$ is the total energy density and $\Omega$ is the sum of matter, radiation and vacuum densities expressed in units of critical density, $\rho_c$, as follows:
\begin{equation}
\Omega \equiv \Omega_m + \Omega_r + \Omega_{\Lambda}, \ \text{where} \  \Omega_i \equiv \rho_i/ \rho_c.
\label{Friedmann_equation}
\end{equation}

\noindent The critical density is defined as $\rho_c \equiv \frac{3H_0^{2}}{8\pi G}$, where $H_0$ is the present value of the Hubble parameter: $H_0 \equiv 100 \ h \ km \ s^{-1} Mpc^{-1}$. The Hubble parameter, in turn, parametrises the expansion rate of the Universe:
\begin{equation}
\frac{H}{H_0} = \left[\Omega_r a^{-4} + \Omega_m a^{-3} + \Omega_k a^{-2} + \Omega_\Lambda \right]^{\frac{1}{2}}.
\label{Friedmann_equation}
\end{equation}

\noindent Cosmological distances contribute to the computation of observables, so we need accurate predictions for those. A first comoving distance is the \emph{comoving radius}, $\chi$. Out to an object at scale factor $a$ (or, equivalently, at redshift $z=(1/a) -1$) it is defined as follows:

\begin{equation}
\chi(a) = \int_{a}^{1} \frac{da'}{a^{2}H(a')}.
\label{chi_a}
\end{equation}
\\
\noindent
Using the comoving radius \textsf{PyCosmo} evaluates the \emph{comoving angular diameter distance}, $r$, as:

\begin{equation}
r(\chi)= 
\begin{cases}
R_0 sinh(\frac{\chi}{R_0}), \ open \\
\chi, \ \ \ \ \ \ \ \ \ \ \ \ \ \ \   flat \\
R_0 sin(\frac{\chi}{R_0}), \ \ \ closed, 
\end{cases}
\end{equation}
\\
\noindent
where $R_0$ is the present value scale radius. The scale radius is defined as $\frac{R}{R_0} = a = (1+z)^{-1}$ and $R_0 = \frac{c}{\kappa H_0}$, where $c$ is the speed of light and $\kappa$ is defined as follows:

\begin{equation}
\kappa^{2}= 
\begin{cases}
1-\Omega, \ open \\
1,  \ \ \ \ \ \ \ flat \\
\Omega - 1, \  closed.
\end{cases}
\end{equation}
\\
\noindent
The \emph{comoving angular diameter distance} is related to the \emph{angular diameter distance}, $D_A$, and the \emph{luminosity distance}, $D_L$, according to $D_A = a^2 D_L = a r(\chi)$. The luminosity distance, in turn, is used to compute the distance modulus, $\mu = 5log_{10}(D_L/pc)-5$.

\subsection{Linear perturbations}
\subsubsection{Growth of perturbations}
\noindent
\textsf{PyCosmo} computes the linear growth factor of matter perturbations, $D(a)$, observing that for sub-horizon modes ($k \gg \eta^{-1}$) and at late times ($a \gg a_{eq}$), we can derive, from the Einstein-Boltzmann equations:

\begin{equation}
\frac{d^{2} \delta_m}{da^{2}} + \left(\frac{dlnH}{da} + \frac{3}{a} \right) \frac{d\delta_m}{da} - \frac{3 \Omega_m H_0^{2}}{2a^{5}H^{2}} \delta_m = 0.
\end{equation}

\noindent Then the growth factor is computed by integrating the differential equation, and  normalised so that $D(a)=a$ in the matter dominated case and $D(a)=1$ when $a=1$. 
Another approach to compute the linear growth factor is implemented in \textsf{PyCosmo} and makes use of hypergeometric functions. This formalism is valid for $\Lambda$CDM only. In Section \ref{validation} we will show the results of the code comparison using both methods.

\subsubsection{Linear matter power spectrum}
\noindent
Theoretical predictions for cosmological observables require knowledge of the matter distribution in the Universe, both at small and large scales. Given the matter density field, $\rho$, we can write it in terms of its mean matter density, $\bar{\rho}(t)$, and the statistical matter density perturbations:

\begin{equation}
 \delta(\textbf{x},t) = \frac{ \rho(\textbf{x},t) - \bar{\rho(t)}}  {\bar{\rho(t)} }.
\end{equation}

\noindent We are interested in the Fourier space overdensity, $\tilde{\delta(\textbf{k})}$, which is the Fourier transform of the density fluctuations. The power spectrum, $P(k)$, is given by the average of overdensities in Fourier-space:
\begin{equation}
\langle \tilde{\delta}(\textbf{k}) \tilde{\delta}(\mathbf{k^{\prime}}) \rangle = (2\pi)^{3} P(k) \delta^{3}(\textbf{k}- \mathbf{k}^{\prime}),
\end{equation}
\noindent where $\delta^{3}$ is the Dirac delta function. \\
\noindent In addition to the Boltzmann Solver solution for the linear power spectrum, other approaches are used, typically based on numerical simulations. In this context approximate functions have been proposed. The fitting functions implemented in \textsf{PyCosmo} for the linear power spectrum are the \textit{Eisenstein $ \&$ Hu}, described in \cite{EH}, and a polynomial fitting function, namely \textit{BBKS} \citep{bbks1997}.

\subsection{Non-linear perturbations}
\noindent As briefly described above, on large scales (small $k$) the power spectrum can be calculated from linear perturbation theory. On small scales, evolving structures in the Universe become non-linear and perturbation theory breaks down. In analogy to the approximate functions for the linear power spectrum, also the non-linear power spectrum can be computed using fitting functions, following the same approach based on numerical simulations. A recently developed method, described in \cite{Bart2016} and \cite{Bart2017}, proposes the prediction of the non-linear power spectrum without using N-body simulations, but through non-perturbative analytical computation. We describe below the two non-linear fitting functions implemented in \textsf{PyCosmo}, \texttt{Halofit} \citep{Halofit_Smith2003, Halofit_Takahashi2012} and the model proposed in \cite{mead2015} and originally implemented in the \texttt{HMCode}. Future code developments will also explore the analytical approach.

\subsubsection{Non-linear power spectrum}
\noindent The \textit{Halo Model} describes the dark matter density field as a superposition of spherically symmetric haloes, with mass function and internal density structure derived from cosmological simulations. The power spectrum can be written as:
\begin{equation}
\label{halomodel}
P(k) = P_{1H}(k) + P_{2H}(k),
\end{equation}

\noindent where $P_{1H}(k)$ and $P_{2H}(k)$ are denoted the \textit{one-halo} and \textit{two-halo term}, respectively. The first relates to the profile of the spherical haloes, while the second accounts for their spatial distribution, considering that their positions are correlated. For more details concerning the \textit{Halo Model} we refer the reader to \cite{Peacock2000,Seljak2000, Cooray2002}. The non-linear fitting functions implemented in \textsf{PyCosmo} are described below.

\paragraph{HaloFit} Predictions for the non-linear matter power spectrum, following the fitting function \texttt{Halofit} \citep{Halofit_Smith2003} and its revisions presented in \cite{Halofit_Takahashi2012}, are both implemented in \textsf{PyCosmo}. Both papers propose the formalism described in eq. \ref{halomodel}, where each term is a parametric function. The revised model provides updated fitting parameters, based on more accurate simulations.

\paragraph{\textit{Mead et al. model}} \textsf{PyCosmo} includes a first \textit{Python} implementation of a revised version of the \textit{Halo Model}, to which we already referred as the \textit{Mead et al. model} \citep{mead2015}, originally implemented in the \texttt{HMCode}. In this model physically-motivated new parameters are added to the \textit{Halo Model} formalism, in particular a smoothing parameter between the \textit{one-halo} and the \textit{two-halo} terms, and further parameters used to describe the effects of baryonic feedback on the power spectrum. The latter are found from a set of high-resolution N-body simulations and from OWLS hydrodynamical simulations which investigate the effect of baryons. As in the original \texttt{HMcode}, three different models accounting for baryons are available: a more general model including prescriptions for gas cooling and heating, star formation and evolution and supernovae feedback, called \texttt{REF}; a model which adds to \texttt{REF} the AGN feedback, called \texttt{AGN}; and a model which is similar to \texttt{REF}, called \texttt{DBLIM}, which includes a more complete treatment of the supernovae feedback, described in \cite{vanDaalen2011}. For more detailed information about these models and how they are defined in the \texttt{HMcode}, we refer the reader to \cite{mead2015}.
In terms of computational speed, part of the \textsf{PyCosmo} code has been implemented in \texttt{cython} to speed up the computations. \textsf{PyCosmo} and the \texttt{HMCode} run at comparable speeds. 

\subsection{Observables }
\subsubsection{Angular power spectrum with the Limber Approximation}
\noindent Many observables in cosmology are expressed in terms of angular correlation functions of random fields, or their spherical harmonic transform, the angular power spectrum. Its calculation gives expressions including several integrals which require numerical evaluation. In order to simplify them, we can use approximation methods, such as the \textit{Limber Approximation} \citep{Limber1953, Kaiser1992, Kaiser1998, Loverde2008}. This prescription is implemented in \textsf{PyCosmo}. In particular, the weak lensing shear power spectrum is expressed as:
\begin{equation}
C_{\ell} = \frac{9}{16} \left( \frac{H_0}{c} \right)^{4} \Omega_m^{2} \int_{0}^{\chi_h} d\chi \left[\frac{g(\chi)}{ar(\chi)} \right]^{2} P \left(\frac{l}{r},\chi  \right),
\end{equation}

\noindent where $\chi$ is the comoving distance and $\chi_h$ the comoving distance to the horizon. $g(\chi)$ is the lensing radial function, which is defined in terms of $p_{\chi}(\chi)$, the probability of finding a galaxy at a comoving distance $\chi$:

\begin{equation}
g(\chi) = 2 \int_{\chi}^{\chi_h} d\chi^{\prime} p_{\chi}(\chi) \frac{r(\chi) r(\chi^{\prime} - \chi)}{r(\chi^{\prime} )},
\end{equation}

\noindent where $p_{\chi}(\chi)$ is normalised as $\int d\chi p_{\chi}(\chi) =1$.  \\
In this work we use the lensing power spectrum, $C_{\ell}^{\gamma \gamma}$, as an example of observable.

\subsubsection{Line-Of-Sight integrals}
\noindent The Boltzmann Solver includes a first python implementation of the $C_{\ell}^{TT}$, using the line-of-sight integration. In this method, described in detail in \cite{LOS1996}, the temperature field is a time integral over the product of a source term and a spherical Bessel function, therefore splitting between the dynamical and geometrical effects on the anisotropies. The source function, which can be computed semi-analytically, is defined as follows:

\begin{equation}
S(k,\eta) = g\left( \Theta_0 + \Psi + \frac{\dot{u}_b}{k} + \frac{\Pi}{4} + \frac{3\ddot{\Pi}}{4k^{2}}\right) + \dot{g}\left(\frac{u_b}{k} + \frac{6\dot{\Pi}}{4k^{2}}\right) + \ddot{g}\left(\frac{3\Pi}{4k^{2}}\right) + e^{- \tau} \left(\dot{\Psi} - \dot{\Phi}\right),
 \end{equation}

\noindent where $g(\eta)$ is the visibility function, defined in terms of the optical depth as $g(\eta) = - \dot{\tau} e^{- \tau}$ . The terms in $\Theta_0 + \Psi$, $u_b$ and $\Pi$ are the Sachs-Wolfe, Doppler and polarization terms, respectively, while the $\dot{\Psi} - \dot{\Phi}$ term describes the Integrated Sachs-Wolfe effect. \\
The temperature field is computed along the line of sight as:

\begin{equation}
\Theta_{\ell}(k,\eta) = \int_{0}^{\eta_0} d\eta S(k,\eta) j_{\ell}[k(\eta_0-\eta)],
\end{equation}

\noindent where $j_{\ell}(\eta)$ is the spherical Bessel function of $\ell$ order. The temperature field, normalized to the density perturbations for dark matter at present time ($\delta_0$), is integrated over the wave-numbers to get the angular power spectrum:

\begin{equation}
C_{\ell}^{TT} = \frac{2}{\pi} \int dk \ k^{2} P(k) \left| \frac{\Theta_{\ell}(k)}{\delta_0(k)}\right|^{2},
\end{equation}

\noindent where $P(k)$ is the linear power spectrum computed at present time.

\section{Implementation}\label{implementation}

\subsection{Architecture}\label{architecture}
\begin{figure}[h!]
	\centering
	\includegraphics[scale=0.3]{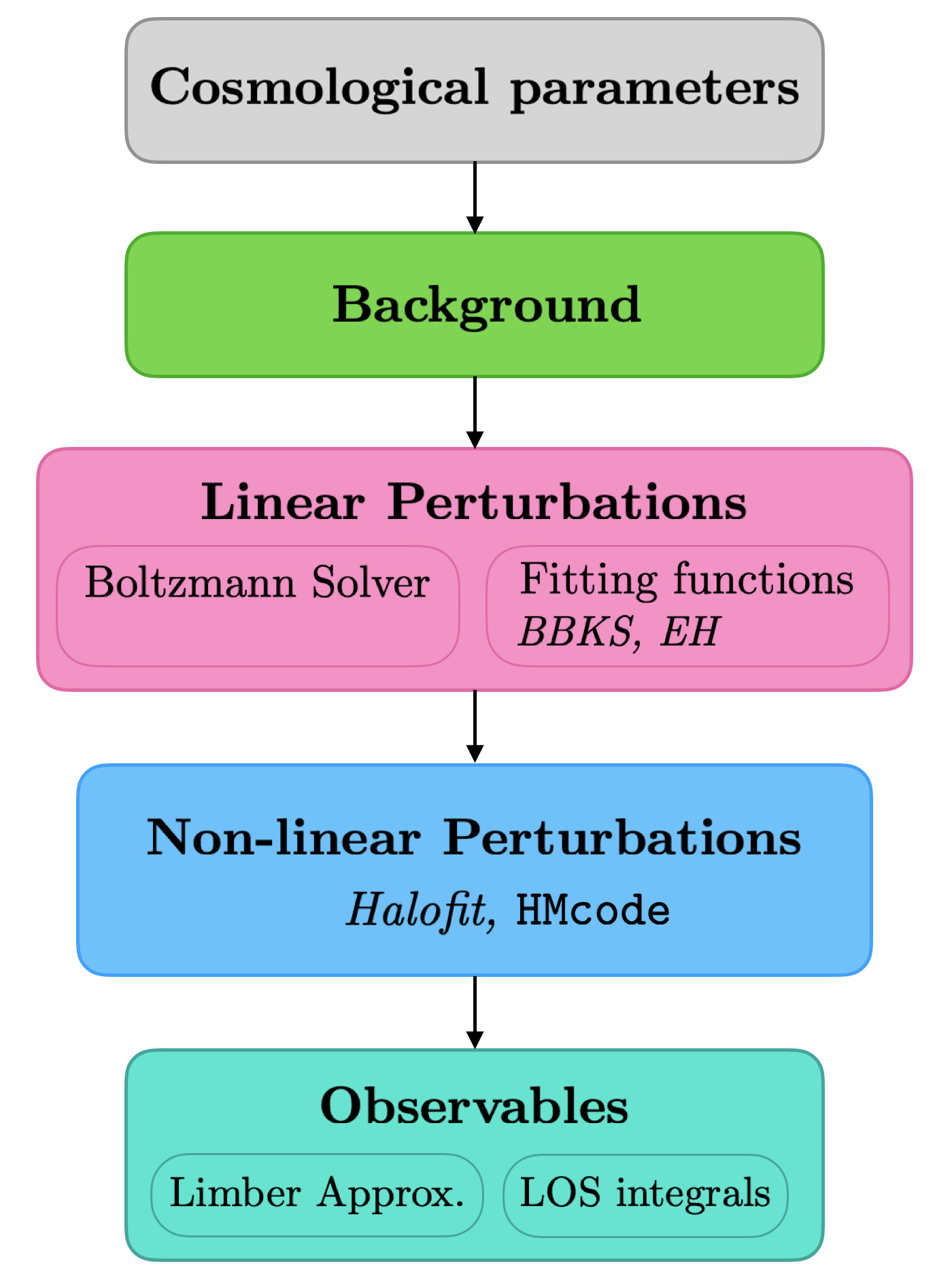}
	\caption{Flow-chart summarizing the \textsf{PyCosmo} architecture. From the top: \textit{Cosmological parameters} refers to the initial cosmological setup, which affects all the computations. The \textit{Background} class computes the Hubble parameter and comoving distances. It is followed by the \textit{Linear Perturbations} and \textit{Non-linear Perturbations} modules, which include various methods to compute matter power spectra (described in Section \ref{theory}). The \textit{Observables} module at the end of the chart calls all the other modules before.}
	\label{architecture_scheme}
\end{figure}

\noindent The flow chart in Figure \ref{architecture_scheme} shows the code architecture. After instantiating \textsf{PyCosmo}, the user can set the cosmology through a set-function which or, equivalently, an internal configuration file. The latter can be modified also to choose the method to compute the matter power spectra. The \textit{Background} class computes basic background quantities, such as the Hubble parameter and comoving distances. The \textit{Linear Perturbations} class provides the linear power spectrum either through the Boltzmann Solver or through fitting functions. The output is then used to compute the non-linear power spectrum in the \textit{Non-linear Perturbations} class. In turn, this module offers a choice of different fitting functions. The power spectrum is involved in computing the observables by the class \textit{Observables}. The theoretical models implemented in this routine are described in Section \ref{theory}.

\subsection{Symbolic calculations}
\noindent As shown in the flow-chart in Figure \ref{architecture_scheme}, one of the classes implemented in 
\textsf{PyCosmo} provides solutions to the set of Einstein-Boltzmann equations, which govern the linear evolution of perturbations in the Universe. The novelty of this solver is its approach to the equations themselves, which are symbolically represented through the \textit{Python} package \textit{Sympy}. The symbolic representation provides the user a convenient interface to manipulate the equations and implement new cosmological models. The equations are then simplified by a \texttt{C++} code generator before being evaluated. For more details about how the solver computes a numerical solution for them, we refer the user to a previous work, \cite{Alex2018}, which focusses on the \textsf{PyCosmo} Boltzmann solver.

\subsection{Unit tests} 
\label{unit_tests}
\noindent Each class shown in Figure \ref{architecture_scheme} is associated with a unit-test routine. It consists in a series of functions testing the methods implemented in each class. These tests perform code-comparison between \textsf{PyCosmo} and the other codes, and check whether the agreement passes a certain numerical accuracy. Every time the code is updated, the developer can check through unit-tests also the impact the new implementations might have on pre-existing parts of the code. The analysis presented later in Section \ref{validation} shows the results of code-comparison which is incorporated in the unit-tests. The \textit{coverage} refers to the amount of code tested and validated in each module through unit-tests. Currently the \textsf{PyCosmo} modules have the following coverage: $100 \%$ for the \textit{Background} class, $91 \%$ for the \textit{Linear Perturbations}, $96 \%$ for the \textit{Non Linear Perturbations} using the \textit{Halofit} fitting function, $95 \%$ for the \textit{Non Linear Perturbations} using the \texttt{HMCode} model and $96 \%$ for the \textit{Observables} class.

\subsection{PyCosmo Hub} 
\begin{figure}[h!]
	\centering
	\includegraphics[scale=0.325]{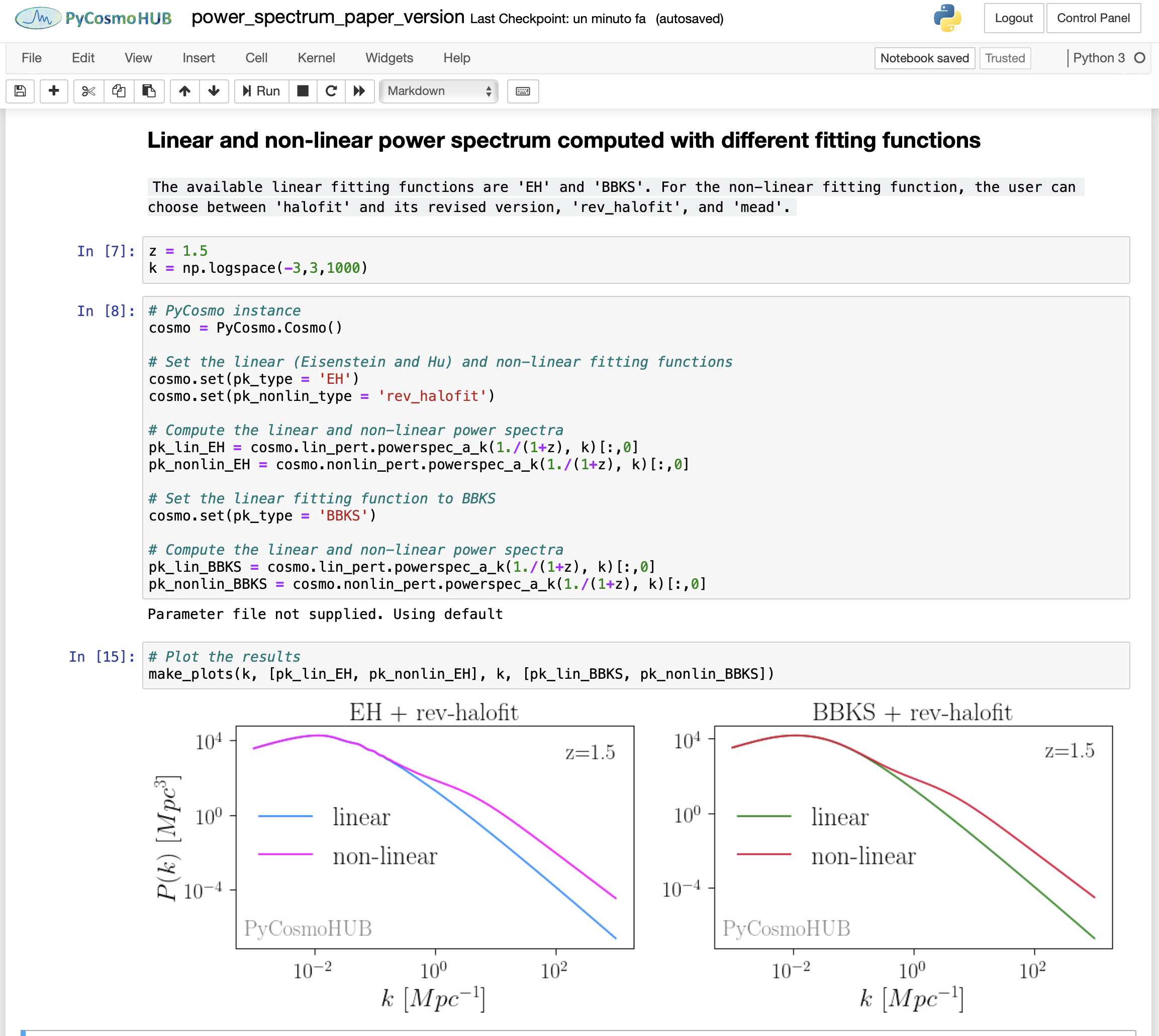}
	\caption{A screenshot of a \texttt{Jupyter Notebook} running on the \textsf{PyCosmo Hub}.}	
	\label{pycosmohublogo}
\end{figure}

\noindent \textsf{PyCosmo} is conceived as a multi-purpose cosmology calculation tool in \textit{Python}, and designed to be interactive and user-friendly. As discussed above, the usage of the \textit{Sympy} package is part of this concept. Indeed, \textsf{PyCosmo} is user-friendly not only in its numerical implementation, but also in terms of its public interface: in order to make its usage immediate to the user, we make \textsf{PyCosmo} publicly available on a hub platform, called \textsf{PyCosmo Hub} (see a screenshot in Fig.\ref{pycosmohublogo}). Its current version, accessible from this link, \url{https://pycosmohub.phys.ethz.ch/hub/login}, includes \texttt{Jupyter} tutorial-notebooks illustrating how to use the code and shows the results of the code-comparison analysis through a series of static notebooks. 
The hub currently hosts the most recent versions of the codes \texttt{CLASS} and \textsf{iCosmo}, which can be run by the users. 
The \textsf{iCosmo} code, originally written in IDL language, is interpreted on the hub through GDL, an open source library alternative to IDL. The $\textsf{PyCosmo}$ version installed on the hub can be downloaded via $pip$. Further information about the code release and documentation is available on this web page: \url{(https://cosmology.ethz.ch/research/software-lab/PyCosmo. html}. The users accessing the hub have space to write their own notebooks, make their own calculations and save the results locally, without the need of installing any software. In this context, the hub is conceived to be useful both for educational purposes and for promoting cosmological inferences in the cloud, in a new dynamic way of teaching and doing research.

\section{Validation and code comparison} \label{validation}
\noindent
In order to assess the level of accuracy in the computation of cosmological observables, \textsf{PyCosmo} monitors its own predictions internally and making comparisons with other cosmology codes. The reliability of every function in \textsf{PyCosmo} is checked through \emph{unit tests}, described in Section \ref{unit_tests}.
In this section, we show the main results from those tests: overall we obtain a good agreement between the codes, both using a fiducial cosmology and testing their response by varying the cosmological setup. We compare the algorithms also in terms of execution speed, with the result that \textsf{PyCosmo} runs at a speed comparable with the other codes. \\

\subsection{Cosmological setup and conventions}\label{settings}
\noindent The tests performed to assess the agreement between the codes are of two kinds, either referring to a fiducial cosmological setup or testing the robustness of the code to changes of cosmological parameters. We assume as our fiducial cosmology: $h = 0.7, \ \Omega_m = 0.3, \  \Omega_b = 0.06, \  \Omega_c = 0.24, \ n_s = 1, \ \sigma_8 = 0.8, \ N_{\rm{eff}} = 3$. We vary cosmology in ranges of $h$ and $\Omega_m$: $h \in [0.4, 0.9]$, $\Omega_m \in [0.2,0.7]$, and we produce heatmaps to show the agreement between the codes across the $(h, \Omega_m)$ parameter space. In this section we include the heatmaps only for the background computations and for the linear and non-linear power spectra, showing those for the other classes in Appendix B.\\

\noindent To illustrate trends as a function of redshift in our fiducial cosmology, for instance in terms of background quantities (cosmological distances, linear growth factor), we consider a redshift range of $z \in [0,9.5)$ with 5000 grid points. If we vary the cosmological parameters, we consider redshift in the range $[0,6)$, maintaining the same number of points. When we compare the non-linear power spectrum to the \texttt{HMcode}, we compute it as a function of wavenumbers, $k$, logarithmically spaced between $10^{-3}$ and $10^{4} Mpc^{-1}$, with a total of 200 points. When we compare the power spectra predicted by different codes we use 200 wavenumbers logarithmically spaced between $10^{-3}$ and $10^{3} Mpc^{-1}$, which is the sampling used by default in \textsf{iCosmo}. Testing the angular power spectrum, we choose a sample of multipoles, $\ell$, linearly spaced between $10$ and $10^{4}$, following also in this case the convention adopted in \textsf{iCosmo}.\\
\noindent In each test, the setup described above is matched between the codes, but there are further parameters which need special care in order to make consistent tests. A detailed description of their configuration is given in Appendix A.\\

\noindent In the next paragraphs, we show the results of the code comparisons. The achieved accuracy is quantified in terms of the relative difference between two compared quantities (i.e. distances, power spectra etc.). Given $Q$ a certain cosmological quantity we consider for comparison between \textsf{PyCosmo} and a code $C$, the accuracy is defined as follows:

\begin{equation}
\label{accuracy}
A = \frac{|Q_{\textsf{PyCosmo}} - Q_C|}{ Q_{\textsf{PyCosmo}} },
\end{equation}

\noindent and it is always reported in logarithmic scale. $A$ is a vector including as many points as the two compared quantities. In the heatmaps summarizing the results when varying cosmology, each cell refers to a particular combination of $(h, \Omega_m)$. It is colour-coded by the base-10 logarithm of the maximum accuracy ($Log[MAX(A)]$) and labelled by the dispersion in accuracy ($\sigma(A)$) obtained for the specific cosmological setup it represents.
We structure our analysis as follows: we start with the background quantities, testing the computation of the cosmological distances. We then proceed with the linear perturbations, discussing the level of agreement reached in terms of the linear growth factor and the linear power spectrum. We move to the non linear perturbations showing the comparisons in terms of the non-linear power spectrum. We conclude with the observables, including the weak lensing and the CMB angular power spectra.
We choose this ordering to emphasize the fact that each step, from the background computations to the linear and non-linear perturbations and up to the observables, influences the accuracy reached in the calculation which comes next. We summarize this procedure and the main results later in Table \ref{summary_table}, which gives an overview of the cosmological quantities which can be computed, the settings used for the comparisons and the level of achieved accuracy.

\begin{figure}[h!]
	\centering
	\includegraphics[width=\textwidth]{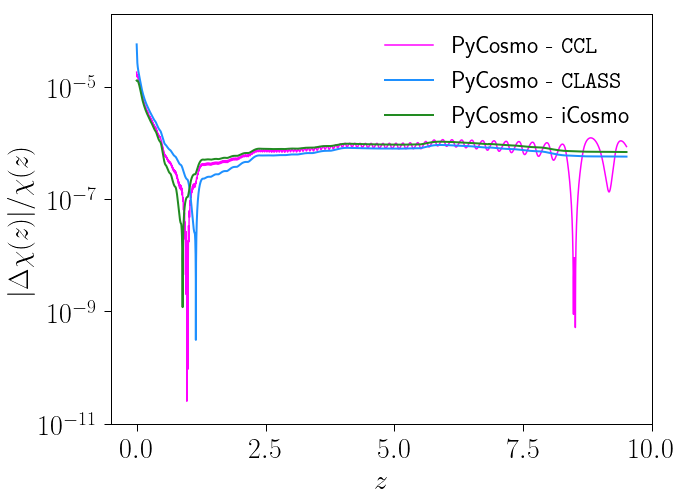}
	\caption{Comparison between \textsf{PyCosmo}, \texttt{CCL}, \texttt{CLASS} and \textsf{iCosmo} in terms of comoving radius, $\chi(z)$, for the assumed fiducial cosmology. The test produces an overall accuracy about $10^{-5}$.}
	\label{chi_a_fiducial}
\end{figure}

\subsection{Background} \label{test_bkg}
\begin{figure}[h!]
	\centering
	\includegraphics[width=\textwidth]{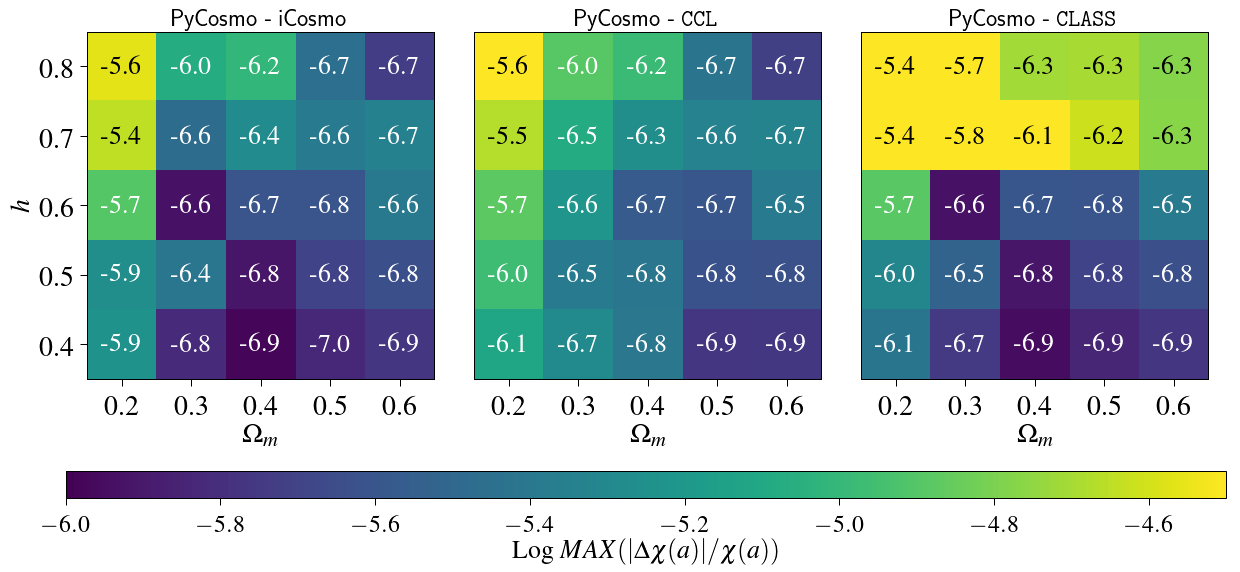}
	\caption{Comparison between \textsf{PyCosmo} and \textsf{iCosmo} (left panel), \textsf{PyCosmo} and \texttt{CCL} (central panel) and \textsf{PyCosmo} and \texttt{classy} (right panel) in terms of comoving radius, $\chi(z)$, for a variety of cosmological parameter values. In the heatmaps each cell refers to a specific combination of $(h, \Omega_m)$. As described in paragraph \ref{settings}, it is color-coded by the maximum accuracy reached in the comparison, and labelled by the dispersion in accuracy. All the results are expressed in logarithmic scale.}
	\label{chi_a_fiducial_vc}
\end{figure}

\noindent
Figure \ref{chi_a_fiducial} summarizes the results of a code comparison made in terms of comoving radius, $\chi$, defined in Eq.\ref{chi_a}. The $y$-axis shows the relative difference between \textsf{PyCosmo} and the other codes, normalised to \textsf{PyCosmo} (see Eq.\ref{accuracy}), as a function of redshift, $z$, up to redshift $z=10$. An overall accuracy around $10^{-6}$ is observed, with oscillations between $10^{-9}$ and $10^{-5}$ at lower redshifts.
We repeat the same test by varying cosmology, as shown in Figure \ref{chi_a_fiducial_vc}. As explained in the paragraph \ref{settings}, the heatmaps are colour-coded by the maximum relative difference occurring between \textsf{PyCosmo} and \textsf{iCosmo} (left panel), \textsf{PyCosmo} and \texttt{CCL} (central panel) and \textsf{PyCosmo} and \texttt{classy} (right panel). Each cell, referring to a combination of $(h, \Omega_m)$, is labelled by the value of dispersion in relative difference obtained for that particular cosmological setup. All the results are expressed in logarithmic scale. Overall we can reach an agreement better than about $10^{-4}$, with small dispersion (up to $\sim 10^{-6}$) overall.

\subsection{Linear Perturbations} \label{lin_pert_results}
\begin{figure}[h!]
	\centering
	\includegraphics[width=\textwidth]{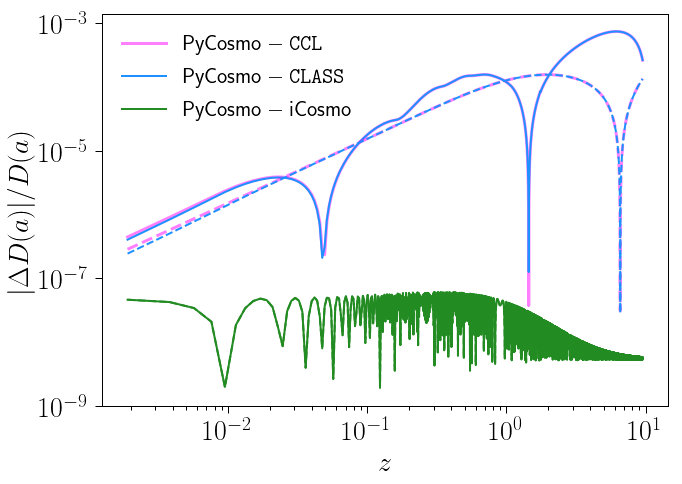}
	\caption{Code comparison in terms of the growth factor, for our fiducial cosmology. The values used for integration accuracy on the ODEINT solvers are specified in detail in Appendix A. \textsf{PyCosmo}, \texttt{CCL} and \texttt{CLASS} agree to better than $10^{-3}$. The lines showing the comparison with \texttt{CCL} and \texttt{CLASS} overlap. The dashed lines show the comparison of the hypergeometric growth factor computed in \textsf{PyCosmo} to the integrated growth factor computed with \textsf{iCosmo}, \texttt{CCL} and \texttt{CLASS}. The dashed and the solid lines for the comparison with \textsf{iCosmo} overlap. The agreement between \textsf{PyCosmo} and \textsf{iCosmo} reaches $10^{-7}$.}
	\label{growth_factor_fiducial}
\end{figure}

\noindent Next, we test the linear perturbations both in terms of the growth factor and the linear power spectrum.
\noindent In Fig.\ref{growth_factor_fiducial} we show the results of the code comparison in terms of the linear growth factor, $D(a)$, computed for our fiducial cosmology and with the same settings described in detail in the paragraph \ref{settings} above. Fig.\ref{heatmap_growth_factor} shows the outcome of the same test, but varying cosmological parameters. All the results are displayed in logarithmic scale. Overall the codes are in agreement, plus we notice a difference between the results obtained by comparing \textsf{PyCosmo} to \textsf{iCosmo} ($10^{-7}$) and \textsf{PyCosmo} to \texttt{CCL} and \texttt{CLASS} ($10^{-3}$). This might be due to the different numerical implementations of the algorithm, which have been discussed already in Section 4.1 of \cite{ccl2019}. As a further test we show the comparison in terms of the hypergeometric growth factor , which offers an analytical reference under the assumption of suppressed radiation. In this test, the dashed lines show the comparison between the hypergeometric growth factor computed in \textsf{PyCosmo} and the integrated growth factor computed with \textsf{iCosmo}, \texttt{CCL} and \texttt{CLASS}. We observe an order of magnitude improvement in the achieved accuracy, as also summarised by the heatmap in Fig.\ref{heatmap_growth_factor_hyper}.\\

\begin{figure}[h!]
	\centering
	\includegraphics[width=\textwidth]{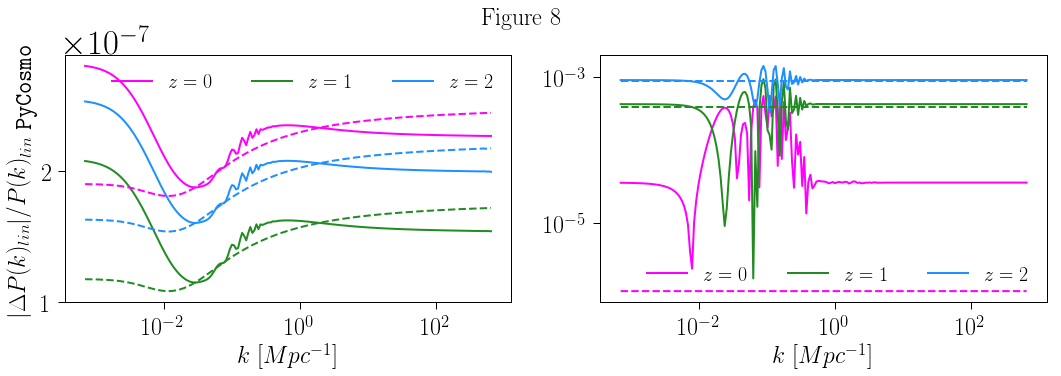}
	\caption{Comparison of \textsf{PyCosmo} with \textsf{iCosmo} (left panel) and \texttt{CCL} (right panel) in terms of linear power spectrum computed with the \textit{EH} (solid lines) and the \textit{BBKS} (dashed lines) fitting functions, for three different redshifts. The $y$-axis on the left panel is not displayed in logarithmic scale for a better visualization.}
	\label{pk_lin_fiducial}
\end{figure}

\begin{figure}[h!]
	\centering
	\includegraphics[scale=0.35]{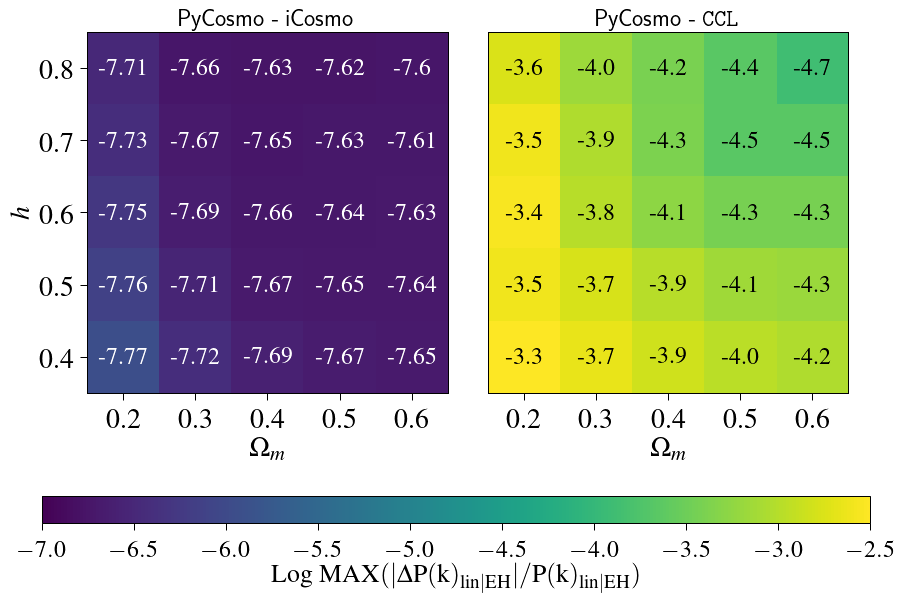}
	\caption{Comparison between \textsf{PyCosmo} and \textsf{iCosmo} (left panel) and between \textsf{PyCosmo} and \texttt{CCL} (right panel) in terms of linear matter power spectrum computed with the \textit{EH} fitting function.}
	\label{pk_lin_heatmap}
\end{figure}

\begin{figure}[h!]
	\centering
	\includegraphics[scale=0.35]{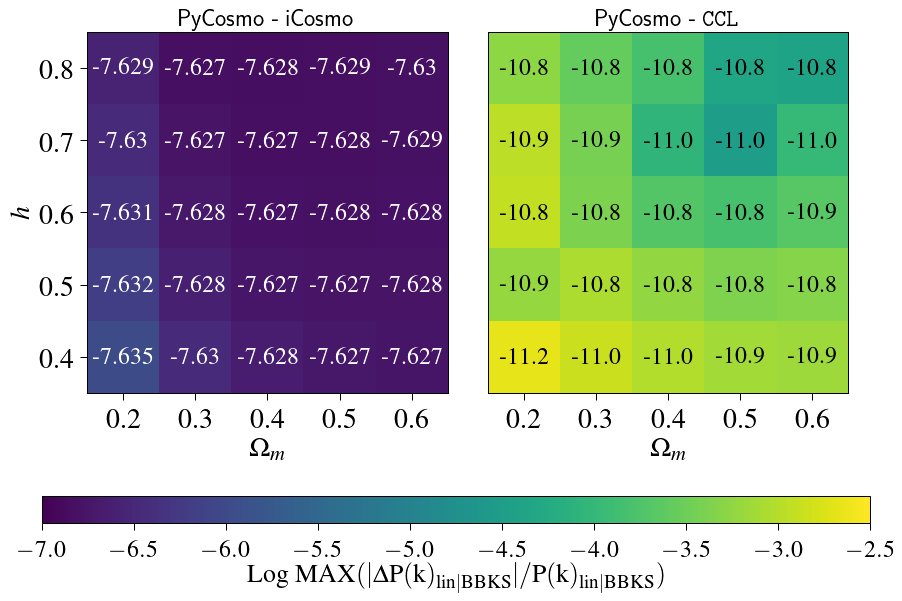}
	\caption{Comparison between \textsf{PyCosmo} and \textsf{iCosmo} (left panel) and between \textsf{PyCosmo} and \texttt{CCL} (right panel) in terms of the linear matter power spectrum computed with the \textit{BBKS} fitting function.}
	\label{pk_lin_heatmap_BBKS}
\end{figure}

\noindent We compute the linear power spectrum both using the \textit{EH} and \textit{BBKS} fitting functions, shown in Fig.\ref{pk_lin_fiducial} with solid and dashed lines, respectively. We compare \textsf{PyCosmo} to \textsf{iCosmo} on the left panel and to \texttt{CCL} on the right panel. 
In both cases the linear power spectrum is computed for our fiducial cosmology and at three different values of redshift, using the same settings described in paragraph \ref{settings}.
Overall we reach a good agreement. The level of accuracy is dominated by the growth factor, whose error propagates into the power spectrum, up to $10^{-7}$ for \texttt{iCosmo} and $10^{-3}$ for \texttt{CCL}, as already shown in Fig.\ref{growth_factor_fiducial}, and increases with time. As observed in the heatmaps in Figures \ref{pk_lin_heatmap} and \ref{pk_lin_heatmap_BBKS}, the same level of accuracy is reached when we vary cosmology. The heatmaps are colour-coded and labelled with the same convention used in Fig.\ref{chi_a_fiducial_vc} and described in paragraph \ref{settings}. 

\begin{figure}[h!]
	\centering
	\includegraphics[width=\textwidth]{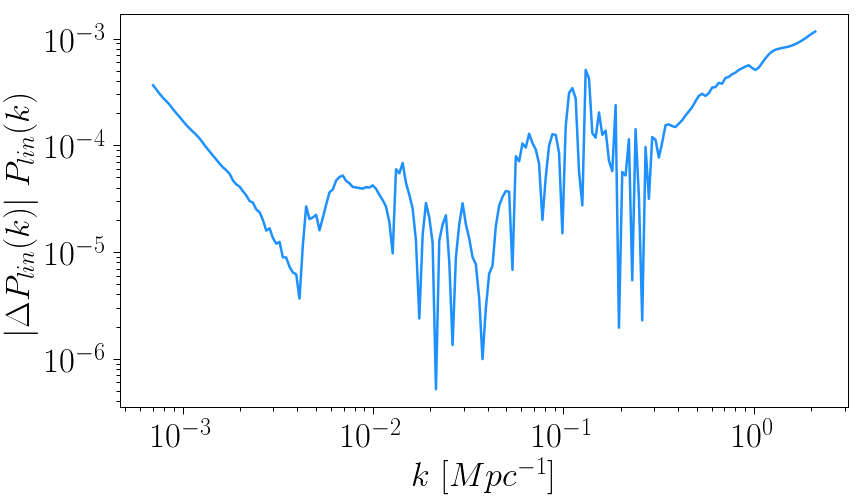}
	\caption{Comparison between \textsf{PyCosmo} and \texttt{classy} in terms of the linear power spectrum computed with the Boltzmann solver. The power spectrum is shown at redshift $z=1$. Both codes were run using high-accuracy settings (described more in detail in Section \ref{lin_pert_results}). A good agreement up to $10^{-3}$ is reached overall.}
	\label{pk_lin_boltz}
\end{figure}

\noindent A good agreement is also observed between \textsf{PyCosmo} and \texttt{classy} when we compare the linear power spectra computed with their respective Boltzmann solvers. Fig.\ref{pk_lin_boltz} shows their relative difference at redshift z=1 for our fiducial cosmology. We ran \texttt{classy} using the same settings listed in the its high-accuracy precision file \texttt{pk\_ref.pre} (available in the public distribution of \texttt{CLASS}), and \textsf{PyCosmo} with $l_{max}=100, \ \epsilon = 3 \cdot 10^{-7}$ and $dt = 10^{-5}$. We reach an agreement better than about $10^{-3}$.

\subsection{Non-linear Perturbations}
\begin{figure}[h!]
	\centering
	\includegraphics[width=\textwidth]{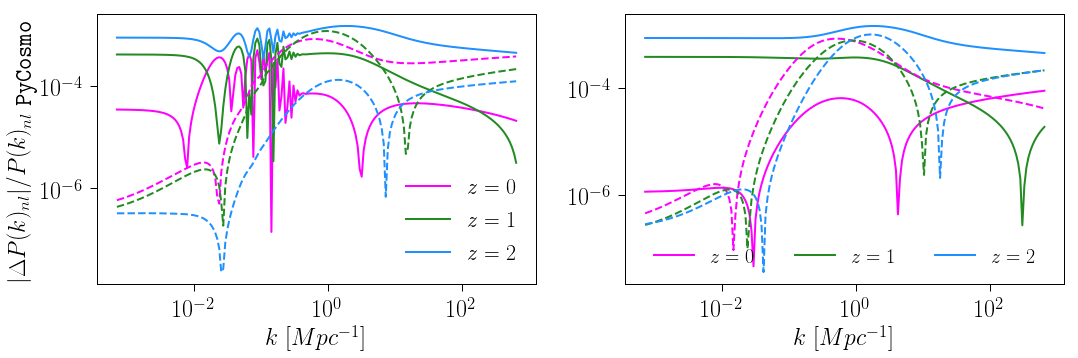}
	\caption{Relative difference in terms of the non-linear matter power spectrum between \textsf{PyCosmo} and \textsf{iCosmo} (dashed lines) and \texttt{CCL} (solid lines). In the comparison between \textsf{PyCosmo} and \textsf{iCosmo} we consider the \textit{Halofit} fitting formula by \cite{Halofit_Smith2003}. The test between \textsf{PyCosmo} and \texttt{CCL} accounts for its revision by \cite{Halofit_Takahashi2012}). The linear fitting formulas used in the computation follow the \textit{EH} and the \textit{BBKS} prescriptions on the left and right panels, respectively.}
	\label{pk_nonlin_fiducial}
\end{figure}

\noindent The accuracy for non-linear perturbations is assessed in terms of the non-linear matter power spectrum and is reported in Fig.\ref{pk_nonlin_fiducial}. In comparing \textsf{PyCosmo} to \textsf{iCosmo} (dashed lines) and to \texttt{CCL} (solid lines), we consider the combinations of non-linear and linear fitting functions which are available in the codes. Therefore we show the following tests:

\begin{itemize}
	\item we compare \textsf{PyCosmo} and \textsf{iCosmo} in terms of non-linear power spectrum computed with the \textit{Halofit} fitting function by \cite{Halofit_Smith2003}. The linear fitting function used is either EH (left panel) or BBKS (right panel).
	\item \textsf{PyCosmo} and \texttt{CCL} are compared in terms of non-linear power spectrum computed with the \textit{Halofit} fitting function by \cite{Halofit_Takahashi2012}. Also in this case, the linear fitting function used is either EH (left panel) or BBKS (right panel).
\end{itemize}

\begin{figure}[h!]
	\centering
	\includegraphics[scale=0.35]{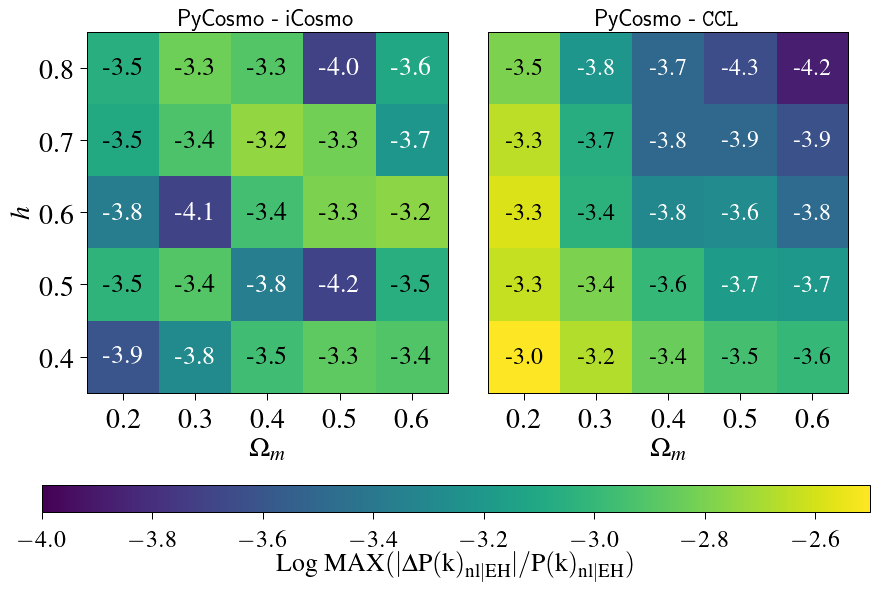}
	\caption{Comparison between \textsf{PyCosmo} and \textsf{iCosmo} (left panel) and between \textsf{PyCosmo} and \texttt{CCL} (right panel) in terms of the non-linear matter power spectrum, computed with \textit{Halofit}+\textit{EH}.}
	\label{heatmap_halofit_EH}
\end{figure}

\noindent We observe that \textsf{PyCosmo} and \textsf{iCosmo} can reach an agreement between $10^{-7}$ and $10^{-4}$. The agreement with \texttt{CCL}, as already observed for the linear power spectrum, is dominated by the growth factor. We obtain analogous results when we vary the cosmological model, as shown in the heatmap of Fig.\ref{heatmap_halofit_EH}: overall the codes are in good agreement, and the algorithm is stable across the parameter space. These observations are valid in both choices of linear fitting functions.\\

\begin{figure}[h!]
	\centering
	\includegraphics[width=\textwidth]{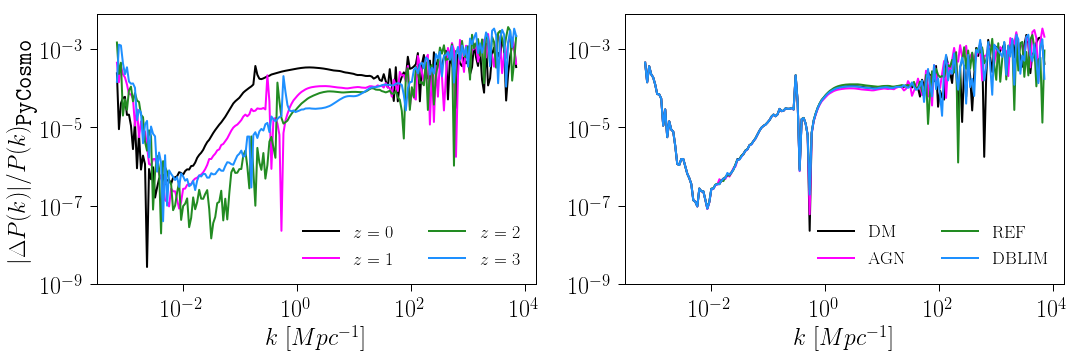}
	\caption{Comparison between the implementations of the \textit{Mead Model} in \textsf{PyCosmo} and in the original code \texttt{HMcode}. Our fiducial cosmology is assumed as the cosmological setup and the non-linear power spectrum is computed assuming the \textit{EH} linear fitting function. On the left panel the comparison is shown for the dark-matter-only case at different redshifts. On the right panel, we add the baryonic feedback for redshift $z=1$. }
	\label{mead_baryons}
\end{figure}

\noindent Moving from \texttt{Halofit} to the \texttt{HMCode}, Fig.\ref{mead_baryons} shows the comparison between its implementation in \textsf{PyCosmo} and the original \texttt{HMcode}, for our fiducial cosmology. The non-linear power spectrum is computed assuming the \textit{EH} linear fitting function. Overall, the computations have been made following the settings described in section \ref{settings}. The left panel is dedicated to the dark-matter-only case and the agreement is studied at different redshifts. The results shown on the right panel take into account the baryonic feedback at redshift $z=1$. In both cases we reach an overall accuracy better than about $10^{-3}$. For more details about the different models of baryonic feedback, we refer the reader to Section \ref{theory}.

\begin{figure}[h!]
	\centering
	\includegraphics[scale=0.35]{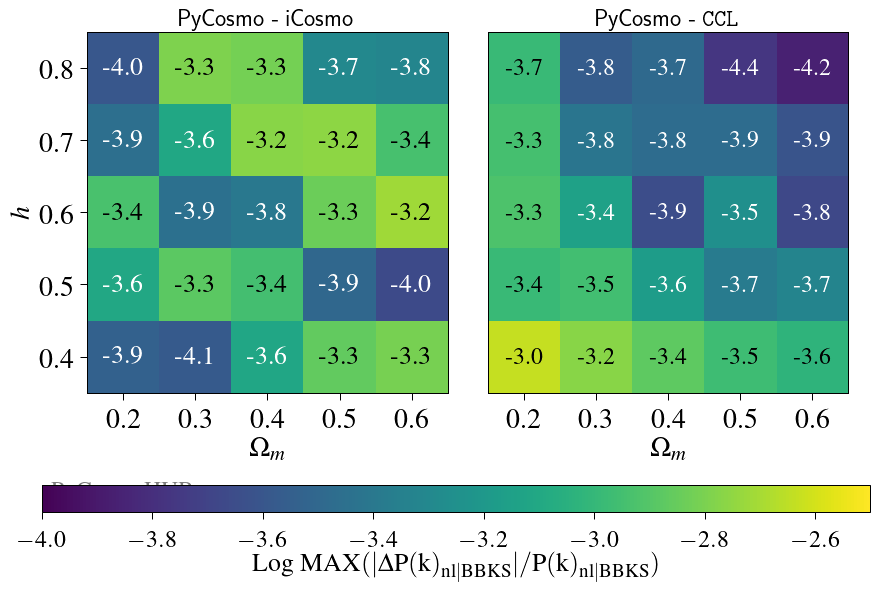}
	\caption{Comparison between \textsf{PyCosmo} and \textsf{iCosmo} (left panel) and between \textsf{PyCosmo} and \texttt{CCL} (right panel) in terms of the non-linear matter power spectrum, computed with \textit{Halofit}+\textit{BBKS}.}
	\label{heatmap_halofit_BBKS}
\end{figure}

\subsection{Observables}
\begin{figure}[h!]
	\centering
	\includegraphics[width=\textwidth]{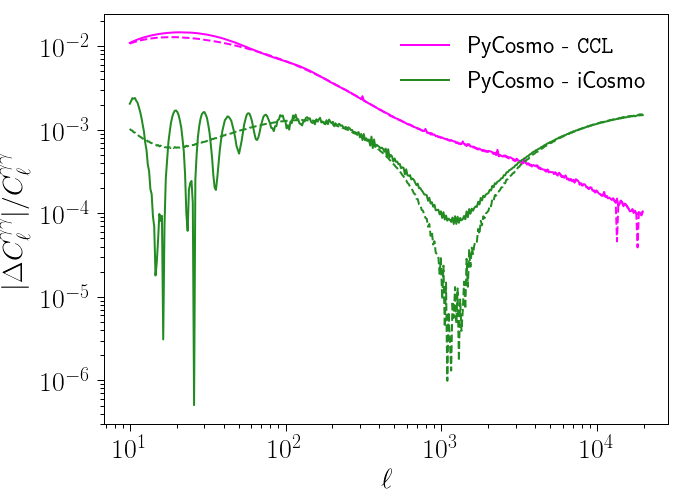}
	\caption{Comparison in terms of $C_{\ell}^{\gamma \gamma}$ between \textsf{PyCosmo} and \textsf{iCosmo} (green lines) and \textsf{PyCosmo} and \texttt{CLL} (magenta lines). \textit{Halofit} and its revised version are used to compute the $C_{\ell}^{\gamma \gamma}$ in the two respective comparisons. \textit{Halofit} is matched both with EH (solid lines) and BBKS (dashed lines) linear fitting functions.}
	\label{cls_nonlin_fid}
\end{figure}

\begin{figure}[h!]
	\centering
	\includegraphics[width=\textwidth]{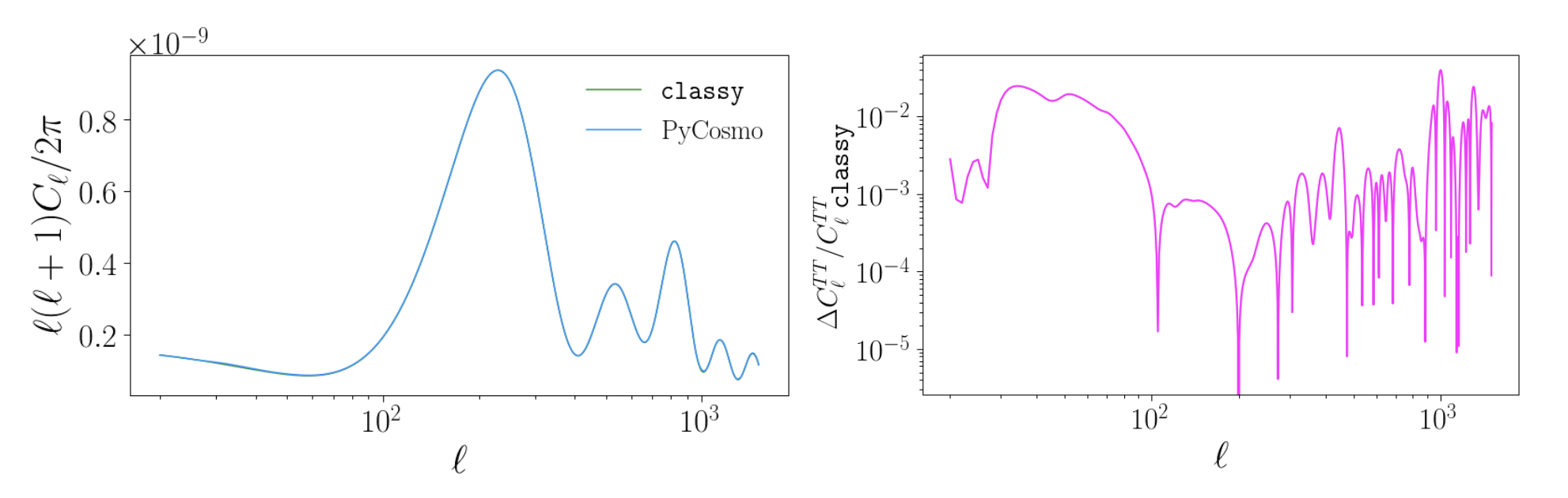}
	\caption{Preliminary CMB angular power spectrum computed with the \textsf{PyCosmo} Boltzmann Solver (on the left) and absolute difference with the same output from \texttt{classy} (left panel). All the terms entering the source function (Sachs-Wolfe, Doppler, Polarization and Integrated Sachs-Wolfe) are considered for the computation in both codes.}
	\label{LOS_cls}
\end{figure}

\noindent We test the observables computed by \textsf{PyCosmo} in terms of the lensing power spectrum ($C_{\ell}^{\gamma \gamma}$) and the CMB angular power spectrum ($C_{\ell}^{TT}$). Figure \ref{cls_nonlin_fid} shows the comparison to \textsf{iCosmo} (green lines) and \texttt{CCL} (magenta lines) for our fiducial cosmology, in terms of $C_{\ell}^{\gamma \gamma}$. The non-linear power spectrum involved in the calculation is computed with the \textit{Halofit} fitting formula, combined with both EH (solid lines) and BBKS (dashed lines) fitting functions. The heatmaps in Figures \ref{heatmap_cls_nonlin_EH} and \ref{heatmap_cls_nonlin_BBKS} show the same test by varying the cosmological parameters. Overall we recover an accuracy up to $\sim 10^{-3}$ for \textsf{iCosmo} and at the percent level with \texttt{CCL}. The heatmap in Fig.\ref{heatmap_cls_linear} shows the comparison between \textsf{PyCosmo} and \texttt{CCL} when the $C_{\ell}^{\gamma \gamma}$ are computed with a linear power spectrum, either using the EH or the BBKS fitting function. Also in this case we reach the same level of accuracy as in the previous test.\\
Fig.\ref{LOS_cls} shows preliminary results from our first Python implementation of the $C_{\ell}^{TT}$ computed with the line of sight integration. The left panel shows the good agreement between the two Boltzmann Solvers, \textsf{PyCosmo} and \texttt{classy}. More details will be reported in a future paper describing the updates and the performance of the new version of the \textsf{PyCosmo} Boltzmann Solver.

\subsection{Summary}\label{table}
\noindent Table \ref{summary_table} represents a summary of the code-comparison described in this paper. It shows the level of agreement between the codes reached in terms of background quantities, power spectra and observables.
Each entry quantifies the agreement using the notation ${\phi_{fid}}_{+ \Delta_{-}}^{- \Delta_{+}} \ (\sigma_{fid})$, which is  explained as follows. \\
We consider a certain observable, $Q(\mathbf{x})$, where $\mathbf{x}$ can be, for instance, a collection of values in redshift or wavenumbers. When we run two different codes we get two independent samples of the same observable, $Q(\mathbf{x})$ and $Q^{\prime}(\mathbf{x})$. For each code we compute their relative difference, expressed as $|Q(\mathbf{x}) - Q^{\prime}(\mathbf{x})| / Q(\mathbf{x})$, and then extract the maximum relative difference, $\phi$, and the dispersion, $\sigma$, of this distribution. 
We repeat the same computations $N$ times, varying cosmological parameters. We get a collection of maximum relative differences, $\Phi = [\phi_1, \phi_2, ..., \phi_{\rm{fid}}, ..., \phi_{N}]$ and the dispersions of their respective distributions, $\Sigma = [\sigma_1, \sigma_2, ..., \sigma_{\rm{fid}}, ..., \sigma_N]$, where $\phi_{\rm{fid}}$ and $\sigma_{\rm{fid}}$ refer to the values obtained for our fiducial cosmology. These values are expressed in logarithm base 10. 
From $\Phi$ we extract $\phi_{\rm{max}}$ and $\phi_{\rm{min}}$, which represent the worst and the best agreement we could obtain by exploring the parameter space of cosmological parameters. In this context we have $|\phi_{\rm{min}}| \ge \phi_{\rm{fid}}, \  |\phi_{\rm{max}}| \le \phi_{\rm{fid}}$. \\
In the notation used in the table $ \Delta_{-}$ and $ \Delta_{+}$ are the distances between $\phi_{\rm{fid}}$ and the worst and best agreement, respectively: $ \Delta_{-} = \phi_{\rm{max}} - \phi_{\rm{fid}}, \  \Delta_{+} = \phi_{\rm{min}} - \phi_{\rm{fid}}$. Therefore the notation ${\phi_{fid}}_{+ \Delta_{-}}^{- \Delta_{+}} \ (\sigma_{fid})$ gives the agreement and the dispersion obtained for our fiducial cosmology, plus the maximum and minimum agreement we get by varying cosmological parameters. \\
All the results reported in the table were obtained following the settings described in section \ref{settings}, at redshift $z=1$. For the comparison with the \texttt{HMcode} we show the maximum and minimum agreement at $z=0$ for the fiducial cosmology, together with the dispersion accuracy for the fiducial cosmology. The same applies to the comparison with the software \texttt{classy} in terms of the CMB angular power spectrum.\\
The hyphenated entries symbolize cases where a certain computation is not available in one of the codes, so no comparison is currently possible.

\begin{table}[H]
	\begin{tabular}{|l|c|c|c|c|}
		\hline
		\multicolumn{1}{|c|}{}            & \textsf{iCosmo}   & \texttt{CCL}   & \texttt{CLASS}   & \texttt{HMCode}           \\ \hline
		\multicolumn{5}{|l|}{\cellcolor[HTML]{aff0ca}\textbf{Background}}                                    \\
		$H(a)$                            &  \textcolor{ForestGreen}{\textbf{-15.2 $ \ $ (-16.1)}}  &  \textcolor{BurntOrange}{\textbf{-4.9 $ \ $ (-5.6)}}  &  \textcolor{ForestGreen}{\textbf{-8.1 $ \ $ (-8.7)}}  & $-$              \\
		$\chi, D_L, D_A(a)$               & \textcolor{ForestGreen}{ $\mathbf{-5.5_{+0.9}^{-0.5} \ (-5.7)}$ } &  \textcolor{ForestGreen}{$\mathbf{-5.1_{+0.6}^{-0.7} \ (-5.6)}$} & \textcolor{ForestGreen}{ $\mathbf{-4.3_{+0.2}^{-1.6} \ (-5.7)}$} & $-$              \\ \hline
		
		\multicolumn{5}{|l|}{\cellcolor[HTML]{FFDAE0}\textbf{Linear Perturbations}}                           \\
		$D(a)$                            & \textcolor{ForestGreen}{$\mathbf{-7.2_{\textcolor{White}{+}0.0}^{-0.1} (-7.9)}$} & \textcolor{ForestGreen}{$\mathbf{-3.5_{+0.7}^{-0.5} (-4.0)}$}& \textcolor{ForestGreen}{$\mathbf{-3.5_{+0.7}^{-0.5} (-3.9)}$} &  $-$               \\
		$D(a)$ hyper                      & \textcolor{ForestGreen}{$\mathbf{-7.2_{\textcolor{White}{+}0.0}^{-0.1} (-7.9)}$} & \textcolor{ForestGreen}{$\mathbf{-3.8_{+0.7}^{-0.5} (-4.4)}$}& \textcolor{ForestGreen}{$\mathbf{-3.8_{+0.7}^{-0.5} (-4.4)}$}&  $-$               \\
		$P(k)_{\rm{lin|BBKS}}$            & \textcolor{ForestGreen}{$\mathbf{-6.8_{+0.8}^{+0.0} (-7.6)}$} &  \textcolor{ForestGreen}{$\mathbf{-3.4_{+0.7}^{-1.1} (-10.9)}$} & $ - $ & $-$              \\
		$P(k)_{\rm{lin|EH}}$              & \textcolor{ForestGreen}{$\mathbf{-6.7_{+0.8}^{+0.0} (-7.7)}$} & \textcolor{ForestGreen}{$\mathbf{-3.0_{+0.6}^{-0.9} (-3.8)}$}& $ - $ & $-$             \\ 
		$P(k)_{\rm{lin|boltz}}$              & $ - $ & $ - $ &  \textcolor{ForestGreen}{$\mathbf{-2.9 \ (-4.1)}$} & $-$             \\ \hline
		
		 \multicolumn{5}{|l|}{\cellcolor[HTML]{A8E1FF}\textbf{Non-linear Perturbations} $P(k)_{\rm{nl|Halofit}}^{(*)}$} \\
		BBKS + S.                         & \textcolor{ForestGreen}{$\mathbf{-3.1_{+0.4}^{-0.6} (-3.6)}$} & $ - $ & $ - $ & $-$               \\
		EH + S.                         & \textcolor{ForestGreen}{$\mathbf{-2.9_{+0.2}^{-0.8} (-3.4)}$} & $ - $& $ - $ & $-$               \\
		BBKS + T.                         & $ - $ & \textcolor{ForestGreen}{$\mathbf{-3.4_{+0.8}^{-0.5} (-3.8)}$} & $ - $ &$ - $                \\
		EH + T.                           & $ - $ & \textcolor{ForestGreen}{$\mathbf{-3.1_{+0.7}^{-0.8} (-3.7)}$}& $ - $ & $ - $                 \\
		\texttt{HMCode} + EH                       & $ - $ & $ - $ & $ - $ &\textcolor{ForestGreen}{$\mathbf{[-7.6, -2.6]}$}   \\
		& $  $ & $   $ & $   $ &\textcolor{ForestGreen}{$\mathbf{(-3.3)}$}                   \\ \hline
		
		\multicolumn{5}{|l|}{\cellcolor[HTML]{CAFFF6}\textbf{Observables} $C_{\ell }^{\gamma \gamma (*)}, \ C_{\ell}^{TT}$}             \\
		$C_{\ell }^{\gamma \gamma}$ \small{BBKS}                              & $-$ & \textcolor{BurntOrange}{$\mathbf{-1.9_{+0.4}^{-0.4} (-2.3)}$} & $-$  & $-$                 \\
 	    $C_{\ell }^{\gamma \gamma}$ \small{EH}                                & $-$ &  \textcolor{BurntOrange}{$\mathbf{-1.8_{+0.3}^{-0.4} (-2.3)}$} &  $-$ & $-$                 \\
		$C_{\ell }^{\gamma \gamma}$ \small{S. + BBKS}                         & \textcolor{BurntOrange}{$\mathbf{-2.8_{+0.2}^{-0.1} (-3.4)}$} & $-$ & $-$  & $-$                \\
		$C_{\ell }^{\gamma \gamma}$ \small{S. + EH}                         & \textcolor{BurntOrange}{$\mathbf{-2.6_{+0.0}^{-0.3} (-3.3)}$}  &  $ - $  &  $-$ & $-$                 \\
		$C_{\ell }^{\gamma \gamma}$ \small{T. + BBKS}                         & $ - $  & \textcolor{BurntOrange}{$\mathbf{-1.9_{+0.4}^{-0.4} (-2.3)}$} & $-$  & $-$                 \\
		$C_{\ell }^{\gamma \gamma}$ \small{T. + EH}                           & $ - $  & \textcolor{BurntOrange}{$\mathbf{-1.8_{+0.3}^{-0.5} (-2.3)}$ } & $-$ & $-$                 \\
		$C_{\ell }^{\gamma \gamma}$ \small{ \texttt{HMCode} + EH }                    & $ - $  &  $ - $  & $-$  & $-$                   \\ 
		$C_{\ell}^{TT}$              & $ - $ & $ - $ &  \textcolor{ForestGreen}{$\mathbf{[-5.7,-2.1](-2.7)}$} & $-$        \\ \hline   
		\multicolumn{5}{|l|}{$(^{*})$ S. = Halofit Smith, T. = Halofit Takahashi}                             \\ \hline

	\end{tabular}
	\caption{Summary of the code-comparison between \textsf{PyCosmo}, \textsf{iCosmo}, \texttt{CCL}, \texttt{CLASS} and \texttt{HMcode}. The structure, from the background computations to the observables, follows the order schematically shown in Fig.\ref{architecture_scheme}, highlighting the fact that the accuracy reached in each module propagates in the next one. Each cell quantifies this accuracy: we explain in detail the adopted notation in paragraph \ref{table}.}
	\label{summary_table}
\end{table}

\section{Conclusions}
\noindent \textsf{PyCosmo} is a recent python-based framework providing solutions for the Einstein-Boltzmann equations and making theoretical predictions for cosmological observables. In this paper, we first discuss its architecture and the implementation of cosmological observables, computed in terms of background quantities, linear and non-linear matter power spectra and angular power spectra (Section \ref{architecture}). In order to asses the accuracy of its predictions, \textsf{PyCosmo} is compared to other codes: (\texttt{CCL}), \texttt{CLASS}, \texttt{HMCode} and \texttt{iCosmo}. Details about the codes and the setup used for the comparisons are given in Sections \ref{introduction} and \ref{settings}. The tests, performed by comparing the output of different and independent codes, and presented in Section \ref{validation}, show that \textsf{PyCosmo} is in good agreement with the other codes over a range of cosmological models. It also includes a first \textit{Python} implementation of the \texttt{HMCode}, which provides an accurate prediction for the non-linear power spectrum which can take into account baryonic effects. We release the currently tested and validated version of \textsf{PyCosmo} (without the Boltzmann Solver) and we make it available on an online platform called \textsf{PyCosmo Hub} (Section \ref{architecture}): \url{https://cosmology.ethz.ch/research/software-lab/PyCosmo.html}. On this hub the users can easily access and use \textsf{PyCosmo} without the need of installing the software locally. In this context, \textsf{PyCosmo} presents an easy and user-friendly interface which is accessible to everyone who wants to compute theoretical predictions for precision cosmology.

\section{Acknowledgements}
\noindent We would like to thank Elisabeth Krause for her constructive discussions about the matter power spectrum and the transfer functions implemented in \texttt{CCL}. We also thank Danielle Leonard, Elisa Chisari and Mustapha Ishak-Boushaki for their useful comments concerning \texttt{CCL}.

\section*{References}

\bibliography{mybibfile.bib}

\appendix
\renewcommand{\thefigure}{A\arabic{figure}}
\setcounter{figure}{0}
\section*{Appendix A}\label{appendixA}

\noindent As mentioned in Section \ref{validation}, the tests performed between the codes require matching those not only in terms of cosmology, bu also considering further parameters which change across the codes. They are set as follows:

\begin{itemize}
	
	\item \textsf{iCosmo} (version 1.2): the agreement between \textsf{PyCosmo} and \textsf{iCosmo} has been tested by setting to zero the radiation density component ($\Omega_{r}=0$), according to the default \textsf{iCosmo} setup. However, even in this configuration the CMB temperature is used in both codes to compute the $EH$ linear fitting function. Therefore we set  $T_{CMB}=2.726 K$, assuming for the CMB temperature the same value used in \textsf{iCosmo} . Concerning the growth factor, $D(a)$, both \textsf{iCosmo} and \textsf{PyCosmo} use the ODEINT solver (\textsf{PyCosmo} uses the \texttt{scipy.integrate.odeint} solver). We find an agreement up to $10^{-7}$ if we set the initial condition at $a=10^{-3}$ and the tolerance parameters as follows:
	\begin{itemize}
		\item \textsf{iCosmo} configuration: integration accuracy set to $10^{-4}$, maximum step size to be attempted by the solver set to $10^{-3}$ and first attempted step size set to $10^{-3}$;
		\item \textsf{PyCosmo} configuration: integration accuracy set to $10^{-9}$ in terms of relative tolerance and to $10^{-12}$ as absolute tolerance. First attempted step size set to $10^{-3}$.
	\end{itemize}
	The computation of non-linear perturbations is tested in terms of the \textit{Halofit} fitting formula proposed in \cite{Halofit_Smith2003}, because this version is the one implemented in \textsf{iCosmo}. \textit{Halofit} is checked both assuming the \textit{EH} and \textit{BBKS} linear fitting functions. The matter power spectrum is then used to compute the lensing power spectrum. For the latter, we use \textsf{iCosmo} at its slower speed, so that a higher accuracy can be reached.
	
	\item \texttt{HMcode} (Git version): analogously to the \textsf{iCosmo} setup, the \texttt{HMcode} suppresses the radiation, so we set \textsf{PyCosmo} accordingly. In the \texttt{HMcode} code the CMB temperature enters the computation of the $EH$ linear fitting function as a hard-wired value, $T_{CMB}=2.728 K$. We set it to this value also in \textsf{PyCosmo}. We match the codes also in terms of the growth factor: in the \texttt{HMcode} the accuracy of the ODEINT solver is set to $10^{-4}$ and the initial condition to $10^{-3}$. We find the highest agreement if we assume for \textsf{PyCosmo} the same configuration used already in the comparison with \textsf{iCosmo} (see the details above). The comparison between \textsf{PyCosmo} and \texttt{HMcode} consists in testing the computation of the non-linear matter power spectrum as prescribed in the \texttt{HMcode} model, both in terms of dark matter only and exploring the effects of the baryons on the power spectrum. The algorithm has been implemented in \texttt{Python} in \textsf{PyCosmo}, and involves the \textit{EH} linear fitting function, according to original prescription in  \texttt{HMcode}.
	
	\item \texttt{CCL} (developer version 1.0.0): the comparison between \textsf{PyCosmo} and \texttt{CCL} requires special care in terms of the growth factor. To achieve the best agreement we set \textsf{PyCosmo} so that the initial condition is at $a=0.1$, the relative and absolute tolerance $10^{-3}$ and $10^{-12}$, respectively, and the first attempted step size $10^{-3}$. In addition to the background quantities, we can compare \textsf{PyCosmo} to \texttt{CCL} also in terms of linear and non-linear power spectra. Using the models available in both codes, we are able to compare the linear power spectrum both with the \textit{EH} and \textit{BBKS} fitting functions, and the non-linear power spectrum with the revised \textit{Halofit} fitting function \citep{Halofit_Takahashi2012}, adopting the two linear fitting functions. The matter power spectra are then involved in the computation of the observables, compared in terms of the lensing power spectrum. 
	
	\item \texttt{CLASS} (version 2.7.1): the agreement between \textsf{PyCosmo} and \texttt{CLASS} has been tested by using the \texttt{CLASS} python wrapper \texttt{classy}. When comparing the linear growth factor, we use for \textsf{PyCosmo} the same setup adopted in the comparison with \texttt{CCL}. Since the linear fitting functions \textit{EH} and \textit{BBKS} are not available in \texttt{CLASS}, we compare the linear power spectra computed with the Boltzmann solver. In this particular test, in order to match the several parameters characterising the two solvers and to achieve the highest possible accuracy, we run the original version of \texttt{CLASS} written in $C$ language.
	
\end{itemize}

\appendix
\renewcommand{\thefigure}{B\arabic{figure}}
\setcounter{figure}{0}
\section*{Appendix B}\label{appendixB}
\noindent In this section we show the heatmaps summarizing the code comparison performed by varying the fiducial cosmological setup. More details about the cosmology assumed and the results are discussed in Section \ref{validation}.  For the description of the quantities shown in the heatmaps, we refer the reader to paragraph \ref{settings} and to Figure \ref{chi_a_fiducial_vc}.

\begin{figure}[h!]
	\centering
	\includegraphics[width=\textwidth]{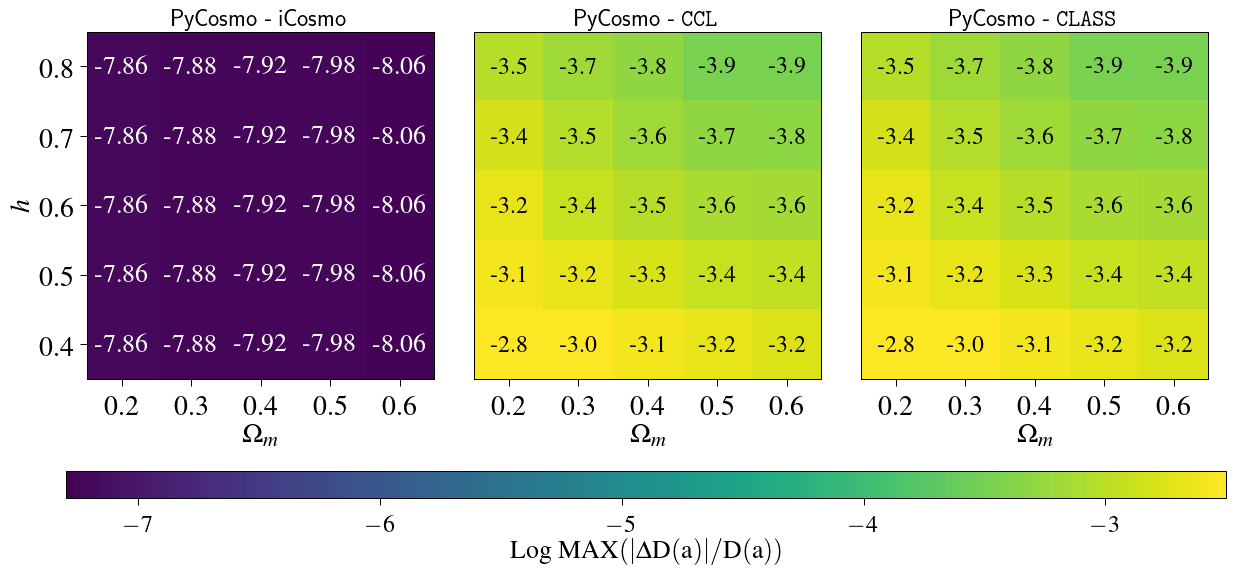}
	\caption{Comparison between \textsf{PyCosmo} and \textsf{iCosmo} (left panel), \textsf{PyCosmo} and \texttt{CCL} (central panel) and \textsf{PyCosmo} and \texttt{CLASS} (right panel) in terms of the linear growth factor.}
	\label{heatmap_growth_factor}
\end{figure}

\begin{figure}[h!]
	\centering
	\includegraphics[scale=0.35]{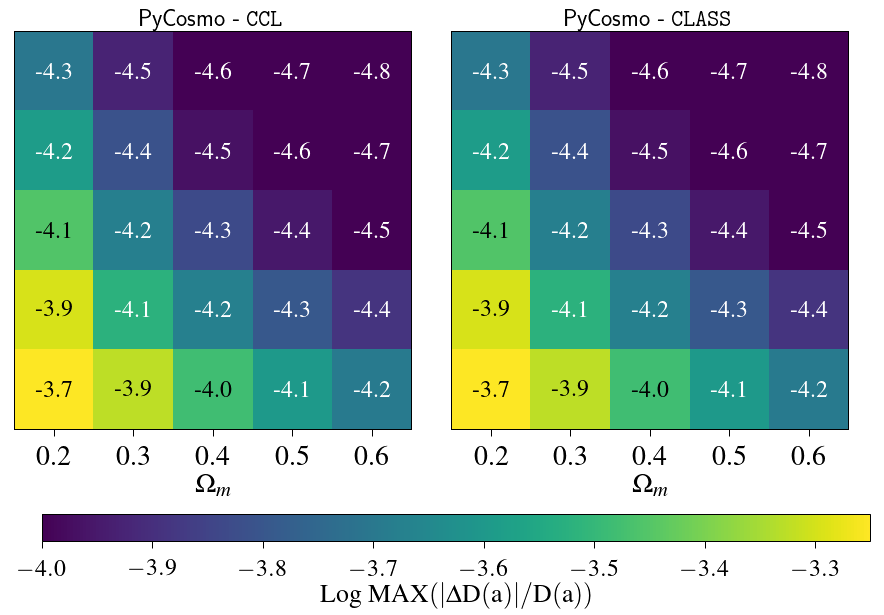}
	\caption{Comparison between \textsf{PyCosmo}, \texttt{CCL} (left panel) and \textsf{PyCosmo} and \texttt{CLASS} (right panel) in terms of the linear growth factor. In this figure the growth factor in \textsf{PyCosmo} is computed with hypergeometric functions.}
	\label{heatmap_growth_factor_hyper}
\end{figure}

\begin{figure}[h!]
	\centering
	\includegraphics[scale=0.35]{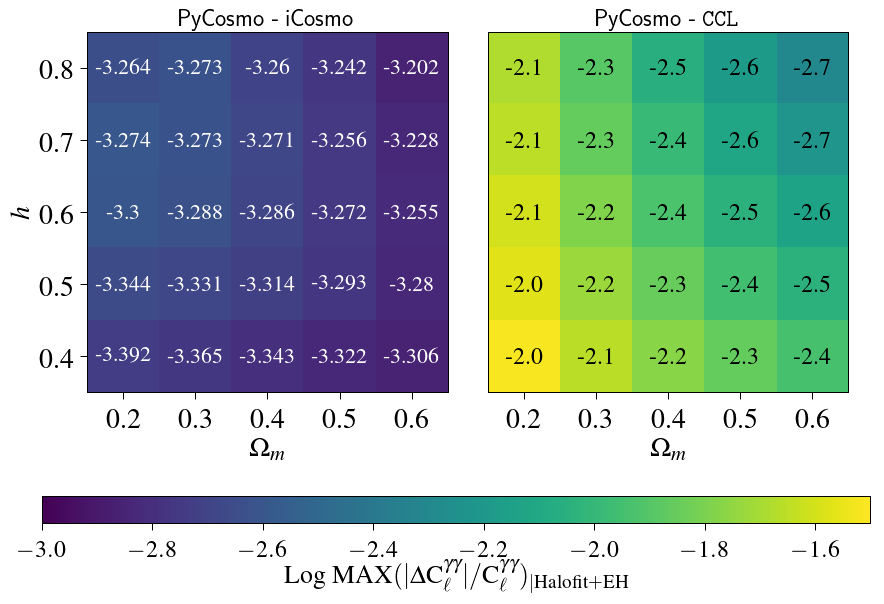}
	\caption{Comparison between \textsf{PyCosmo} and \textsf{iCosmo} (left panel) and between \textsf{PyCosmo} and \texttt{CCL} (right panel) in terms of the lensing power spectrum, computed with \textit{Halofit}+\textit{EH} in the first case and accounting for the revised version of \textit{Halofit} in the second case.}
	\label{heatmap_cls_nonlin_EH}
\end{figure}

\begin{figure}[h!]
	\centering
	\includegraphics[scale=0.35]{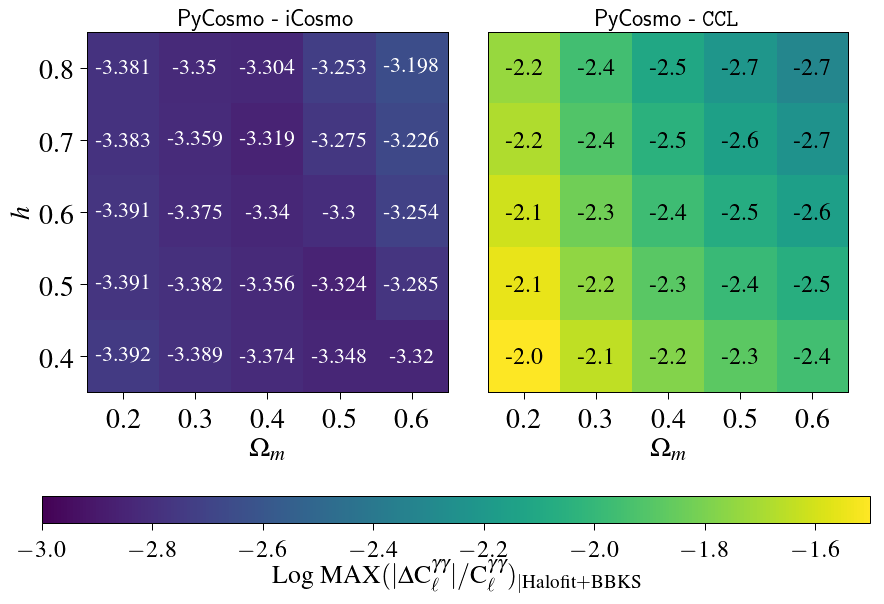}
	\caption{Comparison between \textsf{PyCosmo} and \textsf{iCosmo} (left panel) and between \textsf{PyCosmo} and \texttt{CCL} (right panel) in terms of the lensing power spectrum, computed with \textit{Halofit}+\textit{BBKS} in the first case and accounting for the revised version of \textit{Halofit} in the second case.}
	\label{heatmap_cls_nonlin_BBKS}
\end{figure}

\begin{figure}[h!]
	\centering
	\includegraphics[scale=0.35]{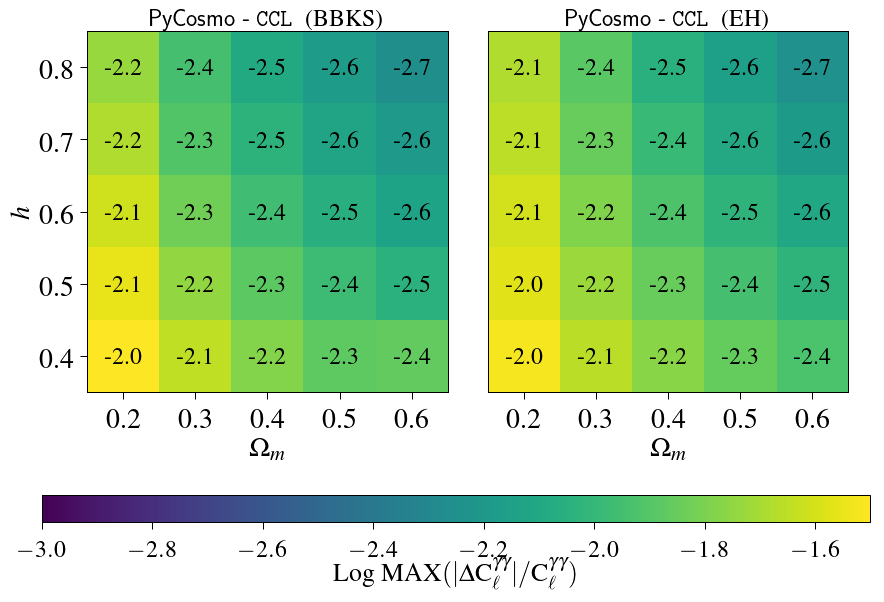}
	\caption{Comparison between \textsf{PyCosmo} and \texttt{CCL} in terms of lensing power spectrum, computed with \textit{BBKS} (left panel) and \textit{EH} (right panel) linear power spectrum. The heatmaps are color-coded by the maximum relative difference between the two compared codes.}
	\label{heatmap_cls_linear}
\end{figure}

\end{document}